\documentclass{epl}

\usepackage{graphicx}
\usepackage{epsfig}
\usepackage{amsmath}
\usepackage{amssymb}

\newcommand{\rmi}{{\rm i}}

\newcommand{\lla}{\left\langle}

\newcommand{\rra}{\right\rangle}

\title{Decoherence in chaotic and integrable systems: \protect\\
A random matrix approach}
\shorttitle{Decoherence: random matrix approach}

\author{T.~Gorin\inst{1}  \and 
T.~H.~Seligman\inst{1,3}}
\institute{
  \inst{1} Centro de Ciencias F\'\i sicas, University of
Mexico (UNAM), Cuernavaca, Mexico\\
  \inst{2} Max Planck Institut f\"ur Kernphysik, Heidelberg,
Germany
}

\begin{document}

\maketitle

\begin{abstract}
In order to analyze the effect of chaos or order on the rate of 
decoherence in a subsystem we aim to distinguish effects of the 
two types of dynamics from those
depending on the choice of the wave packet. To isolate the former we 
introduce a random matrix model that permits to vary the coupling 
strength between the subsystems. The case of strong coupling is 
analyzed in detail, and we find at intermediate times a weak effect of 
spectral correlations that is reminiscent of the correlation hole.
\end{abstract}

New experimental techniques in atomic and quantum optics and more 
recently in solid state physics have made measurements of decoherence
of entangled states possible. The perspective of quantum computing
makes this subject also relevant to applications. In this context the question 
arises how the integrability or chaoticity of the corresponding 
classical systems {\it i.e.} ``quantum chaos''  affects the 
process of decoherence \cite{ref1,ref2,ref3}. Such properties manifest 
themselves both in the spectrum and the wave functions. While the former is 
invariant the latter are basis dependent. Yet that does not mean that the 
latter are irrelevant in a semi-classical context; indeed any wave packet 
localized in phase space will feel special features of the dynamics such as 
KAM tori or short periodic orbits much more strongly, than their
influence on the spectrum. Such effects will be more pronounced 
in integrable or near integrable systems, than in chaotic ones, because 
KAM tori are felt everywhere. If we think of the possible configuration of 
systems this can refer both to the initial wave packet and to the 
Hamiltonian. For the former we can {\it e.g.} think of successive 
laser excitations in Rydberg systems, and for the latter of appropriate 
external fields in atoms or of designed mesoscopic systems. It is 
therefore relevant to ask which properties are due to the nature 
of the system, and which are due to the preparation of the packet.
Studying decoherence we can never do entirely without a packet, but 
to reduce the influence of preparation to a minimum, random matrix 
theory (RMT)  is ideally suited, and we will develop such models for a wide 
range of situations.  

To construct our RMT model we start  from the standard assumption, that
the classical ensembles \cite{ref5} (such as the Gaussian orthogonal 
ensemble (GOE) for time reversal invariant systems) describe the universal 
features induced by classical chaos in quantum systems \cite{ref6}. For the 
classically integrable situation  we expect a random spectrum if we exclude 
harmonic oscillators \cite{ref7}. For the description of a random wave 
packet the orthogonal invariance of the GOE is the ideal tool.
This concept has been extended to the integrable case by the 
introduction of the Poisson orthogonal ensemble (POE) which combines a random
spectrum with orthogonal invariance \cite{ref8}. These models serve 
well to distinguish effects of the invariant properties embodied in 
spectral fluctuations, from others depending on the choice of a specific 
basis or wave packet as well as  on the level density. 

Decoherence will be treated in the framework of unitary time evolution and 
partial traces over subsystems outlined in \cite{ref3}. The model we develop 
allows to analyze various relevant situations, but we limit explicit 
analytic and numeric  calculations in this letter to a strong coupling limit 
in which the preparation of the wave packet and the separation into subsystems 
are completely unrelated to the Hamiltonian. This limit covers some physical 
situations \cite{BSW}, though not the case originally discussed by Zurek 
\cite{ref1}.

To visualize our problem consider a Hamiltonian consisting of three 
terms $H= H^{(0)}+ V^{(1,2)}$ with $H^{(0)} = h^{(1)}+h^{(2)}$ 
where the two terms of $H^{(0)}$ act on different degrees of freedom
of the system, while $V^{(1,2)}$ is an interaction. Note that the $h^{(i)}$ 
may act on one or several degrees of freedom each, and $V^{(1,2)}$ may or may
not induce chaos. Indeed the total system may be integrable and separable 
in a different set of coordinates. Consider {\it e.g.} atoms or molecules 
coupled to the radiative field, spin degrees of freedom coupled to orbital 
ones or coupled cavities with fields.

If we consider this Hamiltonian as a quantum operator we shall denote 
by ${\cal H}_{1}$ and $ {\cal H}_{2}$ the Hilbert spaces on which  
$ h^{(1)}$ and $h^{(2)}$ respectively act. The total Hamiltonian $H$ acts 
on the product space ${\cal H}={\cal H}_{1}\times {\cal H}_{2}$. 
We write the basis states of ${\cal H}_{1}$ as kets with Latin letters such as
$ \vert i \rangle $ and those of ${\cal H}_{2}$ as kets with Greek letters
such as $ \vert \mu \rangle $. The states
$ \vert i, \mu \rangle = \vert i \rangle \vert \mu \rangle $
with indices conveniently written as pairs, form an eigenbasis of $H^{(0)}$.
$H$ is diagonal in a different basis, which we characterize by a single 
index $\alpha$, $\alpha =1\ldots \bar N$. Thus 
$H_{i\mu,i^\prime \mu^\prime} = \sum_\alpha 
O_{i\mu, \alpha}\; E_\alpha\; O^\tau_{\alpha, i^\prime \mu ^\prime}$
where $E_\alpha$ denotes elements of the diagonal energy matrix $E$ and $O$ 
the orthogonal transformation between the two bases.

The Hamiltonians $h^{(1)}$ and $h^{(2)}$ and the interaction term 
may be taken from ensembles or fixed according to the situation we 
wish to analyze. We distinguish between strong and weak interaction.
The interaction strength is usually discussed in terms of the 
spreading width $\Gamma^{(0)}$, which indicates the width of the distribution 
of the expansion coefficients of the eigenstates of $H$ in terms of that of 
$H^{(0)}$. To study the behaviour of an initial state,
that is a product of a state in ${\cal H}_{1}$ with another in ${\cal H}_{2}$,
we have to compare the width $\Gamma$ of this initial state in the eigenbasis 
 with $\Gamma^{(0)}$. Typically we speak of strong coupling if many 
 states are mixed with large amplitudes, and of weak coupling if the  
 states essentially retain their identity with small admixtures of 
 other states which permits a perturbative treatment. In the strong 
 coupling case we can always choose a wave packet such that
  $\Gamma \ll \Gamma^{(0)}$ and we shall include this assumption in 
  the term ``strong coupling'' throughout this paper. In the weak 
  coupling case we always have  $\Gamma \gg \Gamma^{(0)}$ 
  because $\Gamma$ must contain at least four states for a non-trivial 
  decoherence effect. Note that the last condition is 
  not restricted to weak interactions, and intermediate situations
  can occur.

We now construct appropriate matrix ensembles for the three terms in 
the Hamiltonian. The case of weak coupling will not be discussed in detail but
we may mention, that $h^{(2)}$ could describe the central system and $h^{(1)}$
the environment, {\it e.g.} the heat bath, or a finite quality resonator. In
any case both $h^{(1)}$ and $h^{(2)}$ could pertain to either of the above
ensembles, and $V^{(1,2)}$ would typically be symmetric with independent
Gaussian distributed matrix elements. The case studied by Zurek could be
simulated using a GOE with high level density for $h^{(1)}$ 
and a GOE or a POE respectively for $h^{(2)}$. A detailed perturbation study of
weak coupling cases will be published elsewhere.

In the case of strong coupling $h^{(1)}$ and $h^{(2)}$
determine the factor spaces ${\cal H}_{1}$ and $ {\cal H}_{2}$.
Their spectral properties are irrelevant except for their relative spectral 
density in the energy region where the wave packet lives. The interaction 
$V^{(1,2)}$ will be given by the GOE for chaotic systems and by the Poisson 
orthogonal ensemble (POE) \cite{ref8} for integrable ones. Both ensembles are 
given by matrices of the form $O E O^{\tau}$, where $E$ is a diagonal energy 
matrix, and $O$ is a orthogonal matrix distributed according to the Haar 
measure of the orthogonal group. For the GOE the distribution of the energies 
has complicated correlations and a semi circle density, while they are 
independently Gaussian distributed for the POE. The case of strong 
coupling both for integrable and chaotic systems was modeled in 
\cite{BSW} with two-dimensional anharmonic oscillators, and the 
systems considered in \cite{ref3} might be close to this domain.

We expect the spectral 
correlations of the ensembles to represent those of the corresponding 
Hamiltonian systems well. The same is not true for the level densities, which 
are determined by the phase space volume. Yet we are mainly interested in a 
small energy region in which the wave packet is concentrated, and therefore 
we shall assume a constant level density. The effect of rapidly changing
level density  is within the scope of this model. In the above representations 
of our ensembles we therefore replace the energies by unfolded energies with 
the same spectral fluctuations, so that the local average spacing becomes
constant. We shall also use equidistant "picket fence" spectra to complete the 
range of possible spectral correlations. The latter is important, because both 
oscillators in many dimensions and low dimensional systems have spectra 
that are much stiffer \cite{ref9} than what we expect from the 
universal random matrix properties of integrable systems \cite{ref8}. 

Wave packets, while fluctuating in energy space, 
usually have a smooth envelope. We replace it by a sharp cutoff as 
follows: The packet has $n$ components in ${\cal H}_{1}$ and 
$m$ in $ {\cal H}_{2}$, where $n/m$ is determined by the relative spectral
density of the two subsystems in the energy region we consider, and 
$n\times m = N$ is determined by the width $\Gamma = Nd$, where $d$ is the
local mean level spacing. We consider the Hilbert space made up of
the tensor product of the spaces spanned by these vectors
 leading to a $N < \bar N$ dimensional total space. This
 approximation is drastic, but it is important to note that we only 
 restrict the shape of the packet in energy space and {\it not} the 
 fluctuations. The way we truncate the space is consistent 
 with the original structure, in that the truncated packet is again
 a product of functions in the two subspaces, and applying
 the same notation we used for the infinite dimensional subspaces
 to the finite dimensional ones, we still have 
${\cal H}={\cal H}_{1}\times {\cal H}_{2}$.

Consider $\rho_{i\mu, i^\prime \mu ^\prime}$ to be a density matrix 
constructed from a ``product state'' {\it i.e.} is pure with respect to both 
pairs of indices. With other words we have 
${\rm Tr}_1[{\rm Tr}_2 (\rho)]^2 = {\rm Tr}_2[{\rm Tr}_1(\rho)]^2=1$.
Here ${\rm Tr}_1$ indicates the trace with respect to the first (Latin) 
index and ${\rm Tr}_2$ the one over the second (Greek) index. We shall choose 
a density matrix $\rho=\rho(0)$ fulfilling this condition as initial state, 
watch its time evolution and finally obtain ensemble averaged properties
of partial traces. The simplest quantity to analyze is the purity defined as
\begin{equation}
I(t) ={\rm Tr}_1[{\rm Tr}_2 (\rho(t  ))]^2 \ =\  
{\rm Tr}_2[{\rm Tr}_1(\rho(t  ))]^2.
\end{equation}
We are free to interchange summations as all sums are finite. The definition 
of the purity $I$ is related to the idempotency defect or linear entropy,
defined as $1-I$ \cite{ref3}.

Denoting by $ \Delta $ the diagonal matrix with entries 
$\Delta_\alpha =\exp[it  \, E_\alpha]$ we find in the basis of double indices
$\rho(t  )= O\Delta O^\tau   \rho(0) O \Delta ^* O^\tau$ and by consequence
\begin{equation}
I(t  )= {\rm Tr}_1[{\rm Tr}_2(O\Delta O^\tau \rho(0) O 
\Delta ^* O^\tau ) {\rm Tr}_2(O
\Delta  O^\tau    \rho(0) O
\Delta^{*} O^\tau  )]
\end{equation}
To form the ensemble average of this quantity we use that the measure 
factorizes. We take the averages involving energies and the averages involving 
states separately for different terms of the sum. The only object that will not 
be averaged over is the original pure density matrix $\rho(0)$. 
In principle we can perform the averages and then use the idempotency 
condition, but as we average over all orthogonal transformations we 
can reinterpret the states $ \vert i \rangle \vert \mu \rangle $ as 
pertaining to a basis where $\rho(0)$ is diagonal with $\rho(0)_{11,11}=1$
and all other matrix elements zero; it is easy to verify that this is 
possible within the product basis. Using this form of $\rho(0)$ we obtain 
for the two averages:
\begin{align}
 AE_{\alpha,\beta ;\gamma ,\delta} &=
\langle \Delta_{\alpha} \Delta_{\gamma}
 \Delta ^{*}_{\beta} \Delta ^{*}_{\delta}\rangle 
= \langle {\rm exp}[\rmi t(E_\alpha +E_ \gamma - E_{\beta} -E_{\delta)} 
 \rangle \\
AO_{\alpha,\beta ;\gamma,\delta} &=\sum_{\mu,\nu,i,j}
 \lla O_{i\mu,\alpha} O_{11,\alpha}
O_{11,\beta} O_{j\mu,\beta}
O_{j\nu,\gamma} O_{11,\gamma}
O_{11,\delta} O_{i\nu,\delta}\rra 
\end{align}
The averages are connected only because one may force indices to be equal and 
thus reduce the other to a special case; as we shall see below five different 
terms exist.

We will not need all averages over monomials of eight matrix elements
of the orthogonal group, as we concentrate on the relevant time scales.
The ones we will need can be evaluated with Ullah's two-vector formula, if we 
correct an error therein \cite{ref10}. We therefore consider the possible time 
scales in our problem, namely the inverse of the width $\Gamma$ of the packet 
in energy space and the Heisenberg time $1/d$ where $d$ is the  mean level 
spacing.  We have to inspect the effect of these quantities when folded on a 
circle by the  exponential ${\exp}(\rmi t\,E_\alpha)$. We obtain four time 
scales for the evolution of the wave packet:
\begin{itemize}
\item[1)]{
Short times where $t\ll 2 \pi/\Gamma$, and thus perturbation theory 
applies. We will find the expected $t^2$ dependence with a factor given
to leading order by the spreading width.}
\item[2)]{
First filling of the unit circle, where $t=2\pi /\Gamma$, and we will
find a quadratic minimum for the purity with value $(2\pi /\Gamma)= 
1/n+1/m+0(1/N)$.}
\item[3)]{
Long times, where the spectrum winds many times around the unit circle
as $1/d\gg t\gg 2 \pi/\Gamma $, the winding acts as a random number 
generator and eliminates correlations to yield a result 
similar to the one for 2) though sub-leading terms may be different.}
\item[4)]{
Poincar\'e recurrence time $t=2 \pi/d$, where a picket fence spectrum 
will cause exact recurrence, while even for GOE type correlations the
recurrence is essentially wiped out. Yet for low-dimensional systems with 
their long-range stiffness \cite{ref9} and for models
involving harmonic oscillators this part may well be important. The same 
holds if the wave packet is extremely narrow.}
\end{itemize}
We now proceed to derive and illustrate the results for the above. We evaluate 
the averages $AE$ and $AO$ starting with the former. There we deal with four 
indices and the result does not depend on the values of the indices. It 
depends only on whether certain indices are equal or not. If two indices of
the energies coincide, we get either $0$ if they have opposite signs or twice
the energy of their signs are equal. It may readily be seen that five terms 
are possible: 
\begin{alignat}{2}
S_1(t) &= \lla \exp[{-\rmi t \; (E_1-E_2+E_3-E_4)}]\rra &
&= f^4(t) \notag\\
S_2(t) &= \lla \exp[{-\rmi t \; (E_1-E_2)}]\rra &
&= f^2(t) \notag\\
S_3(t) &= \lla \exp[{-\rmi t \; (2E_1-E_2-E_3)}]\rra & 
&= f(2t) f^2(t) \label{Soft}\\
S_4(t) &= \lla \exp[{-\rmi t \; 0}]\rra & 
&= 1 \notag\\
S_5(t) &= \lla \exp[{-2\rmi t \; (E_1-E_2)}]\rra & 
&= f^2(2t) \notag
\end{alignat}
The result for each term corresponds to uncorrelated energies (POE). 
Here $f(t) = \sin(\Gamma t)/(\Gamma t)$ is the Fourier transform of 
the level density, which we assumed to be constant. For GOE spectra the 
evaluation is more difficult, but some general considerations hold 
for any kind of spectrum.  
For long times all terms except $ S_{4} $ go to zero. For short times, on 
the other hand, $ S_{1} $ dominates because it has the largest weight.
We now consider the four time regimes: 

In the short time limit we expand the exponential. 
Due to the symmetry of the energy distribution linear terms in 
$t$ vanish and the quadratic ones survive. These are of two types. 
Each exponential associated with a given index has a quadratic term, and 
indices in the linear terms of two exponentials may coincide. This implies 
that we only need the well-known averages over monomials of fourth order
in the group elements to obtain
\begin{equation}
I(t) = 1-2 \langle E_{\alpha}^2 \rangle \, t^2 [1-(n+m+1)/(N+2)] \; . 
\end{equation}
In the last factor we seem to have a $1/N$ correction. Yet if $n$ and $m$ grow 
as $\sqrt N$ the correction is of order $1/\sqrt N$. If one of the two
dimensions is kept constant, the other becomes proportional to $N$, and 
the second term even more important. Terms resulting from correlations
of the energies are truly of order $1/N$ and were omitted.

The next time scale is that of the first filling of the unit circle,
for which the minimum of the function $f^2(t)$ is reached. We have a 
complicated interplay of different terms and it seems that we need $AO$
completely. To avoid this we can use a trick to obtain the answer at time 
$2\pi/\Gamma$. For uniform density of the spectrum the energy eigenvalues 
multiplied by $2\pi/\Gamma$ are essentially the eigenphases of a circular 
ensemble. For the case of GOE fluctuations the corresponding ensemble is known 
as the circular orthogonal ensemble (COE) \cite{ref5}, which is the ensemble 
of unitary symmetric $N \times N$ matrices $S$. This ensemble has a  unique 
invariant measure. The only approximation we make is that we miss the 
correlations among the first and last levels that exist in a circular ensemble.
In terms of $S$ we obtain
\begin{equation}
I(2\,\pi/\Gamma) = \lla Tr_1[Tr_2(S \,  \rho(0)\, S^*)Tr_2(S\, \rho(0)\, S^*)
\rra \; .
\end{equation}
where $S_{i\mu,j\nu} =\sum_\alpha O_{i\mu,\alpha}
{\rm exp}[\rmi E_\alpha (2\pi/\Gamma)] 
O_{j\nu,\alpha}$. The ensemble average is thus given in terms of averages over 
four symmetric unitary COE matrices, two of which are complex conjugate. 
Such averages are calculated in ref.~\cite{ref11} and we obtain
\begin{equation}
I(2\pi/\Gamma) =  \frac{(n+m) N^2 + \left[3(n+m)+2\right] N - 2(n+m-1)}
{N(N+1)(N+3)} \; .
\end{equation}
As we shall see below this is slightly lower than the long time limit, while 
for the integrable (POE) case the long time limit and the value at 
$t=(2\pi/\Gamma)$ coincide.

In the long time limit $\Gamma t\gg 2\pi$, the process of stretching and 
taking modulo $2\pi$ is a reasonably efficient randomizer for a fluctuating 
set of numbers with correlations such as a GOE spectrum. Therefore the 
eigen-phases on this time scale are random both for the GOE and the POE.
Thus only the term $S_4$ survives, where the indices of energies in conjugate 
terms coincide. The energy dependence, and therefore the time dependence, 
drops out and we are left with averages over the orthogonal group, $AO$. Only
two-vector terms {\it i.e.} averages over elements from two rows of the matrix 
occur and we find
\begin{equation}
I_\infty = \frac{(n+m)N^3 +3[4(n+m)+3]N^2+[35(n+m)+57]N + 48}
{(N+1)(N+2)(N+4)(N+6)} \; .
\end{equation}
For the POE this result holds equally  at time $t=2\pi/\Gamma$, which we have 
discussed above, though it will differ for times near this one. 

While the short-time behaviour is independent of the spectral 
statistics, at the first minimum we find a difference
between GOE and POE. We may ask, what 
happens for very stiff spectra such as the picket fence. 
In this case all correlations are known and we could proceed 
to calculate I(t), but as we never expect an exact picket fence,
we shall give qualitative comments and include
the corresponding results in the numerical calculations shown in the 
figures. We expect the 
first minimum to be at least as deep as for GOE statistics, but 
as in this case the formal limit is nearly reached it cannot be very 
different. On the other hand we do not obtain randomization so  
the purity, after a few oscillations, 
remains at this level until we reach Poincar\'e recurrence.
For times of the order of the Heisenberg time we expect a very different 
behaviour; more precisely if  $t=2\pi/d$. In this case we have exact 
revival in the case of the picket fence, while we certainly expect nothing 
for the random spectra. In the case of GOE fluctuations what we have to 
consider is the width of the $k$th neighbour spacing distribution. 
It is known to increase logarithmically. For $k=1$ we have a width of 
$\approx 1.25$ and for $k=8$ it is already $\approx 1.85$. We thus will see no  
signature in this case.  We should though note two facts: first we will find 
for a picket fence an additional partial revival at half the time mentioned, 
because of the terms with $2t$ in the time evolution; second and more 
importantly the long-range stiffness of spectra in low-dimensional systems 
\cite{ref9} implies a saturation in the width of the $k$th neighbour spacing 
distribution and could therefore show recurrences. This point will not be 
addressed in the present letter.

\begin{figure}
\twoimages[scale=0.35]{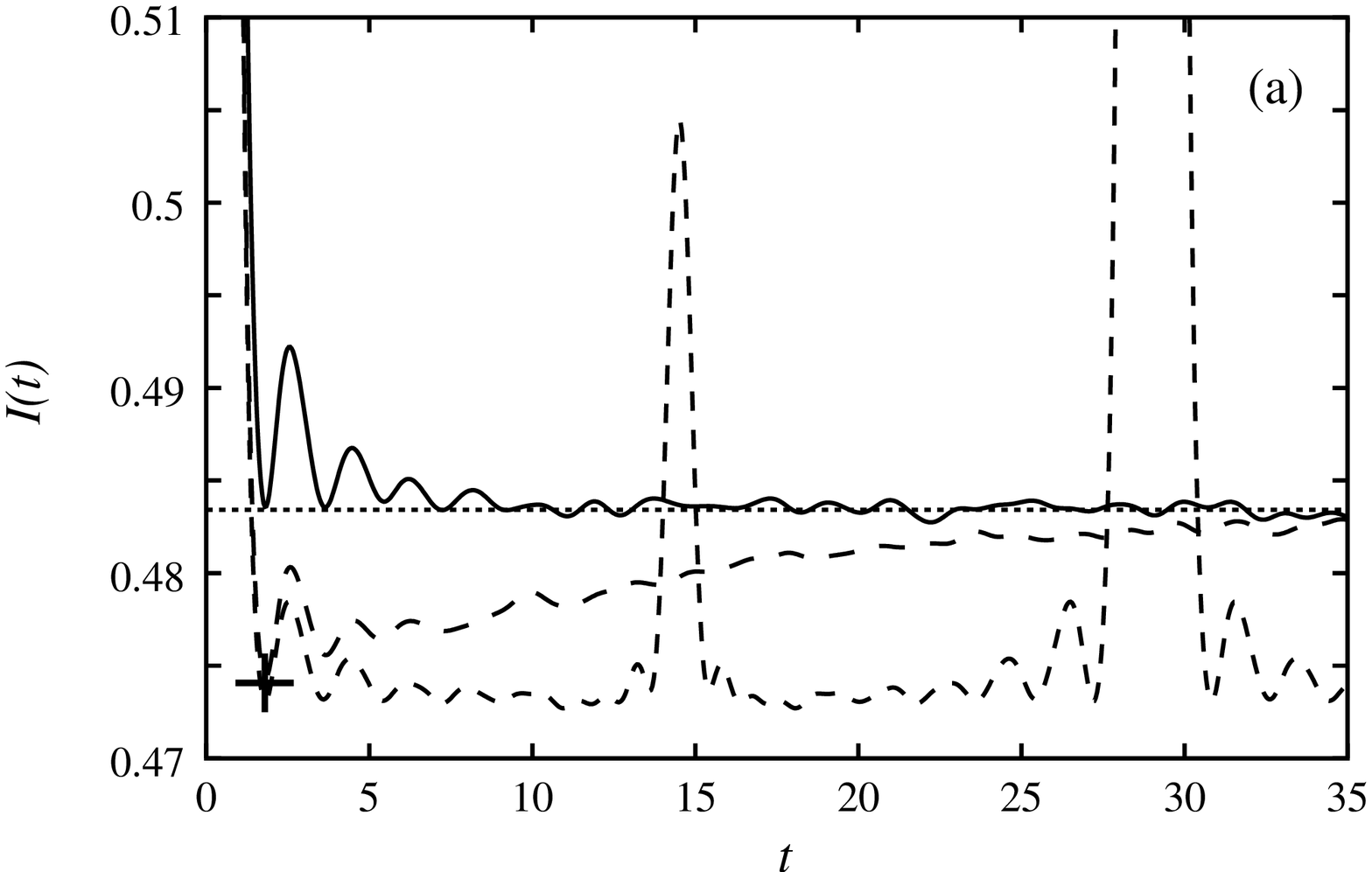}{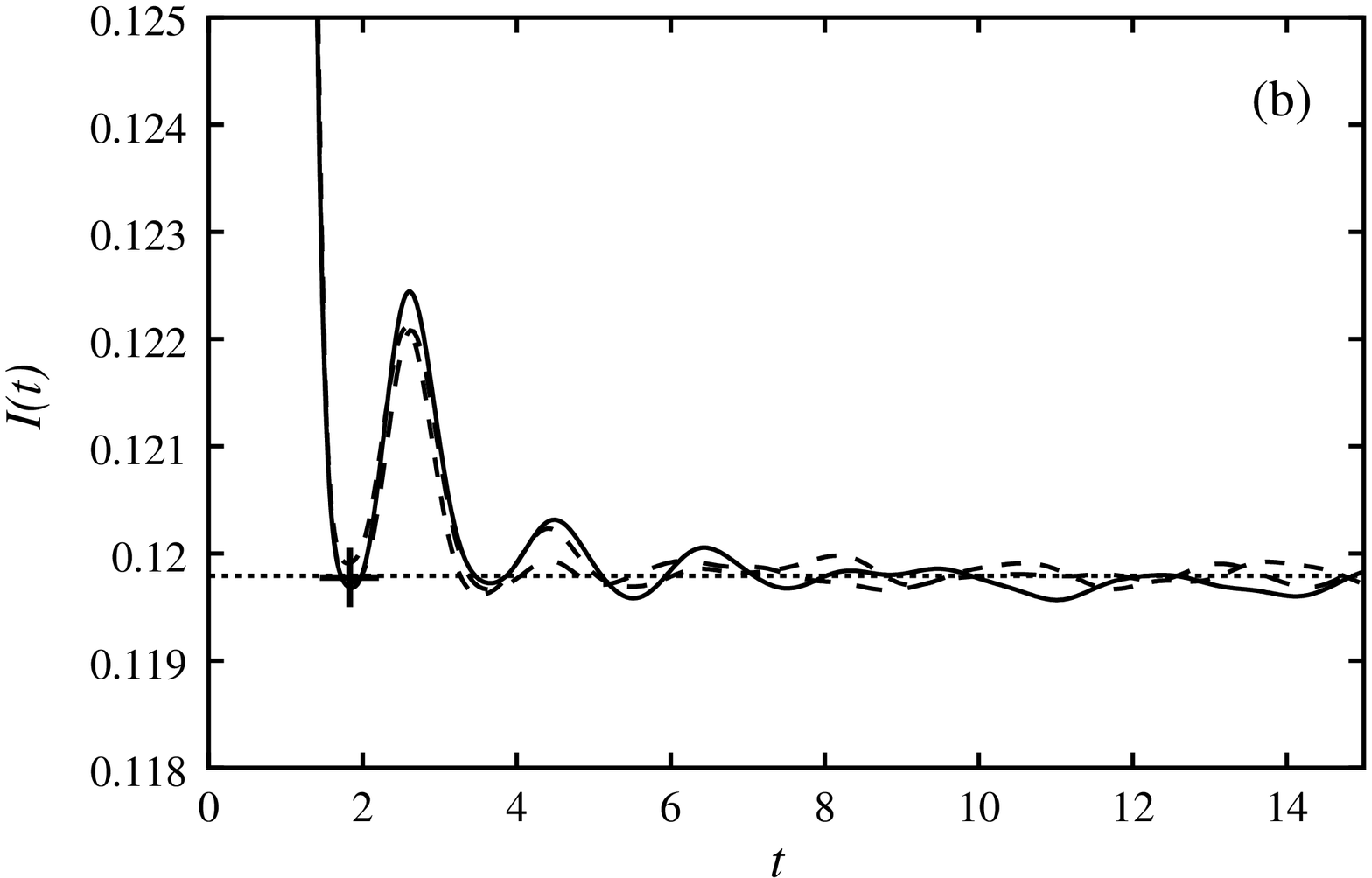}
\caption{The purity $I(t)$ is shown for $m=n=4$ in (a) and $m=10, n=50$ in
(b), starting from an initially pure state as a function of time. 
The ensembles considered are: POE
(solid line), GOE (long dashed line), picket fence (short dashed line). The
value $I_\infty$ (dotted horizontal line). The value at the first minimum
$I(2\pi/\Gamma)$ (upright cross). We suppress the evolution
at short times to see the difference between GOE and POE.}
\end{figure}
 
We illustrate the time dependence discussed by the 
evolution of the purity for the cases $m=4,\,n=4$ and $m=10,\,n=50$, in the
Figures~1(a) and (b) respectively. In both cases we show the results for 
spectra with random and GOE like fluctuations as well as picket fence spectra,
using the same value of $\Gamma$ in all cases.
We notice that our expectations are well fulfilled.
At the beginning all curves are equal, and the first minimum occurs at the 
same time. In Fig.~1a for $N=16$ the minimum is lower for GOE like 
fluctuations and for Picket fence spectra than for random spectra, though 
the effect is quite small. This difference is negligible in Fig. 1b. with 
$N=500$. For large times the GOE- and the Poisson-like fluctuations yield the 
same result even for small $N$. The results at all times coincide with our 
theoretical expectations. This is also true for the recurrences, 
which are only seen for picket fence spectra both at $t=2\pi/d$ and more 
weakly at $t=\pi/d$.

In Fig. 1a the rise of purity after the first minimum follows roughly that of 
the Fourier transform of the two-point function with appropriate scaling. This
is not surprising, because we may expect that a cluster expansion of the 
correlations relevant for the difference from the random case
is dominated by the two-point function. A detailed calculation will be 
reserved for a later publication.

 We have proposed a random matrix approach to decoherence, which 
 allows to isolate the effects of spectral statistics, {\it i.e.}
 the universal features of chaos and integrability on this process.
 We find the behaviour of the purity at short and at 
 intermediate times is dominated by the width of the wave packet in 
 energy space, which is the one property of the packet we have to 
 take into account even in our models. It determines both the 
 prefactor of the initial quadratic behaviour and the position of the 
 first minimum. The plateau reached at long times is
 dominated by the number of levels of the smaller subsystem involved, 
 but the corrections resulting from the number of levels of the 
 larger system are also important. Both are again depending on the width
 of the packet, but in this case the ratio of these numbers is decisive.
 We find correction terms of order
 $1/N$ or larger, that are important for narrow wave packets.
 While the plateau does not depend on integrability or chaos,
 the intermediate time scales show a weak effect of chaos
 that is reminiscent of the correlation hole.
 
 For the strong coupling case we conclude, that the influence of
 spectral statistics on decoherence is  
  small and limited to packets with few eigenstates.
 This implies a  very limited direct impact of chaos
 on decoherence. Yet for specific choices of the wave packets 
 this situation may be very different, if these relate to the 
 invariant tori of the integrable system in Hilbert space, as we know
 from the behaviour of intensities \cite{BSW}.
 For chaotic systems strong deviations from this theory may still 
 occur if the packet had some special relation to features known
 as scars, that may be associated with only
 slightly unstable periodic orbits, parabolic manifolds, 
 or bifurcations that influence the chaotic dynamics at larger scales.
 Due to the mixed phase space used in \cite{ref3} a comparison with the 
 calculations presented there is delicate, but the general picture is 
 not incompatible. Calculations in other entirely chaotic and 
 integrable systems are under way.

\acknowledgments
We thank M.C Nemes and H.A. Weidenm\"uller for stimulating discussions 
and criticism as well as the A. v. Humboldt Foundation, the DGAPA (UNAM) 
project IN112200  and CONACyT project 25192E for financial support.

\end{document}